\documentclass[pdflatex,sn-basic]{sn-jnl}
\usepackage{graphicx}%
\usepackage{multirow}%
\usepackage{amsmath,amssymb,amsfonts}%
\usepackage{amsthm}%
\usepackage{mathrsfs}%
\usepackage[title]{appendix}%
\usepackage{xcolor}%
\usepackage{textcomp}%
\usepackage{manyfoot}%
\usepackage{booktabs}%
\usepackage{algorithm}%
\usepackage{algorithmicx}%
\usepackage{algpseudocode}%
\usepackage{listings}%
\usepackage{lineno}

\usepackage[framemethod=tikz]{mdframed}
\newmdenv[linecolor=black,backgroundcolor=white]{algo}

\definecolor{darkblue}{rgb}{0.0, 0.2, 0.6}
\definecolor{darkgreen}{rgb}{0.0, 0.6, 0.2}
\definecolor{darkred}{rgb}{0.5, 0, 0}

\begin{document}

\title[Identifying atmospheric fronts based on diabatic processes using the dynamic state index (DSI)]{Identifying atmospheric fronts based on diabatic processes using the dynamic state index (DSI)}

\author*[1,2]{\fnm{Laura} \sur{Mack}}\email{laura.mack@geo.uio.no}
\author[2,3]{\fnm{Annette} \sur{Rudolph}}
\author[2]{\fnm{Peter} \sur{Névir}}

\affil[1]{\orgdiv{Department of Geosciences}, \orgname{University of Oslo}, \city{Oslo}, \country{Norway}}

\affil[2]{\orgdiv{Institute of Meteorology}, \orgname{Freie Universität Berlin}, \state{Berlin}, \country{Germany}}

\affil[3]{\orgdiv{Institute of Landscape Architecture and Environmental Planning}, \orgname{Technical University Berlin}, \state{Berlin}, \country{Germany}}

\abstract{Atmospheric fronts are associated with precipitation and strong diabatic processes. Therefore, detecting fronts objectively from reanalyses is a prerequisite for the long-term study of their weather impacts. For this purpose, several algorithms exist, e.g., based on the thermic front parameter (TFP) or the F diagnostic that combines relative vorticity and horizontal temperature gradient. It is shown that both methods have problems to identify weak warm fronts since they are characterized by low baroclinicity. To avoid this inaccuracy, a new algorithm is developed that considers fronts as deviation from an adiabatic and steady state. These deviations can be accurately measured using the dynamic state index (DSI). The DSI shows a coherent dipole structure along fronts and is strongly correlated with precipitation sums. Using the DSI, a new front detection algorithm is developed (called DSI method), which allows to clearly identify the global storm track regions.  The properties of the identified fronts depend on the applied front detection method, whereby fronts identified with the DSI method have particularly high specific humidity. Using a simple estimate for front speed, it is shown that also the front speed depends on the front detection method and that fronts identified using the DSI method have a higher front speed than fronts identified with the TFP method. This can be attributed to the dipole structure of the DSI and thus demonstrates the potential of the DSI to inherently indicate the movement speed and direction in atmospheric flows.}


\keywords{Objective Front Identification, Front Detection, Front Climatology, Diabatic Processes, Dynamic State Index}

\maketitle

\section{Introduction}\label{sec1}
Fronts are air mass boundaries, that occur as far-reaching features from extra-tropical cyclones \citep{Catto2013} and are associated with weather extremes, such as strong precipitation, wind gusts or compound extreme events \citep{Catto2012,Catto2013,Clark2018,RavehRubin2019}. \citet{Catto2013} showed that globally 51\,\% of precipitation extremes occur along fronts and in storm track regions even 90\,\%. 
With climate change, frontal activity is expected to shift poleward and to decrease in the northern hemisphere accompanied by a decrease in front-associated precipitation \citep{Berry2011,Catto2014,Dagon2022}. 
The general increase in specific humidity in a warmer atmosphere is thereby altered non-linearly and spatially inhomogeneously by atmospheric dynamics \citep{Pfahl2017}. \\
Fronts separate different air masses \citep{Bjerknes1922} and manifest themselves as band-like hyper-baroclinic zones \citep{Renard1965}, which initially form due to geostrophic deformation \citep{Hoskins1982}. Since the front-parallel component falls within the synoptic-scale, while the front-perpendicular component falls within the meso-scale (i.e. relative vorticity and Coriolis parameter are of the same magnitude), fronts are a cross-scale phenomenon and theoretically described in a semi-geostrophic Eulerian framework \citep{Hoskins1982}. The coupling between temperature and wind field induces a cross-frontal ageostrophic secondary circulation, which further amplifies the horizontal temperature gradient \citep{Sawyer1956,Eliassen1962} and causes a pronounced area of minimum pressure and maximum relative vorticity along the front \citep{Hoskins1982}. The development of fronts can be diagnostically analyzed with a frontogenesis function \citep{Petterssen1936} or studied based on the isentropic slope tendency that accounts for the opposing effects of diabatic heating and tilting of isentropic surfaces \citep{Papritz2015}. Alternatively, a Lagrangian view of fronts allows to consider different conveyor belts as coherent air streams around cyclones \citep{Wernli1997}. The complex dynamical processes along fronts cause various scale interactions, with larger-scale cyclones and meso-scale convection. \\
The relation between fronts and cyclones is subject of a cause-and-effect discussion with two points of view \citep{Schemm2018}: According to the Bergen school \citep{Bjerknes1922} cyclones develop at the polar front when it becomes unstable, such that northward moving warm air forms a warm front downstream and southward moving cold air a cold front upstream. This type of cyclone is called initial-front cyclone and usually forms over the western oceanic boundary currents \citep{Schemm2018}. The theory of baroclinic instability explains the formation of cyclones from a broad hyper-baroclinic zone \citep{Charney1947,Eady1949}, and fronts can then form as consequence of the cyclogenesis process. These cyclones -- referred to as late-front cyclones -- predominantly form orographically in the lee of large mountain ridges \citep{Schemm2018}. \\
In particular cold fronts play an important role in the initiation of convective cells \citep{Markowski2010}, which was recently systematically studied in a cell-front distance framework by \citet{Pacey2023}. Attributing weather events to fronts thus requires to distinguish their contribution from cyclones and meso-scale phenomena \citep{Rudeva2015}, which calls for automated front detection procedures \citep{Hewson1998}. \\
Due to this complex multi-scale nature of frontal dynamics, diagnostic data-based studies that analyze climatologies and characteristics of fronts statistically from reanalyses or climate projections are of central importance \citep[e.g.][]{Berry2011,Simmonds2011,Catto2013,Catto2014,Parfitt2017,Bochenek2021,Dagon2022,Niebler2022,Pacey2023} -- and the basis for these studies are methods for front identification.
\citet{Hewson1998} pointed out the advantages of automated and reproducible algorithms for front identification, which are referred to as 'objective' methods, compared to hand-drawn front lines on weather maps, which are referred to as 'subjective' methods \citep{Uccellini1992}. 
According to \citet{Hewson1998}, 'objective' methods should be simple, intelligible, accurate, tunable and portable. But even for these 'objective' methods, suitable meteorological variables and threshold values must first be found. Due to the diverse properties of fronts, which cannot always be unambiguously assigned to fronts only \citep{Thomas2019}, various criteria can be designed for this purpose that focus on different characteristics of fronts. \\
\citet{Renard1965} consider fronts as a hyper-baroclinic zone with a pronounced thermal gradient that separates two air masses of less baroclinicity and thus motivates them to introduce the thermal front parameter (TFP) as gradient of horizontal baroclinicity to detect atmospheric fronts threshold-based from gridded data \citep{Hewson1998}. The thermal front parameter can be calculated based on different thermal quantities, such as temperature \citep{Serreze2001}, potential temperature \citep{Berry2011}, equivalent potential temperature \citep{Schemm2015} or wet-bulb potential temperature \citep{Catto2014} (of which the last two also take humidity into account). \citet{Simmonds2011} introduced a front identification method based on a dynamical quantity only, by detecting surface fronts based on a 10 m-wind shift. \citet{Parfitt2017} combines the horizontal temperature gradient as thermal quantity with the relative vorticity as dynamical quantity to one detection variable, called F diagnostic. However, these different detection methods lead to differences in global front climatologies, their seasonal variability and the attribution of precipitation to them \citep{Hope2014,Schemm2015,Soster2022}, which makes it difficult to draw generally valid and reproducible conclusions on these questions. \\
This is why we provide a new alternative view on front detection: We consider fronts as deviation from a stationary, abiabatic and inviscid basic state. This basic state can be exactly derived from the primitive equations and the deviation from it is measured with the dynamic state index (DSI) \citep{Nevir2004,Weber2008}. In the basic state the DSI equals zero, while along fronts strong deviations occur, which are related to ongoing unsteady and diabatic processes associated with latent heat release and precipitation \citep{Claussnitzer2008}. \\
Our aim is to apply the DSI to front detection by developing a new DSI-based front detection algorithm ("DSI method") and compare it in both, a case study and a global climatology, with existing front detection methods and discuss differences in the properties of the detected fronts. With this new DSI method, our study contributes to the discussion about front detection methods, which are the fundamental basis for diagnostic studies investigating the properties, dynamics, scale interaction and weather impact of fronts. \\
The study starts with recapitulating two existing front identification methods (utilizing TFP and F diagnostic, Sec. \ref{chap:frontid}) based on a case study. Then, the DSI is introduced and it is shown how the DSI reflects fronts. Based on this, the new front identification method (''DSI method'') is developed (Sec. \ref{chap:dsi}). Sec. \ref{chap:comparison} compares the three methods based on a global climatology and their annual cycle using ERA5, and examines which properties the detected fronts have.


    \section{Data and Methods}\label{chap:frontid}

    \subsection{Approaches for identifying atmospheric fronts}
    Since fronts separate air masses with different thermal properties, they are characterized by a strong horizontal temperature gradient. In the classical Bergen cyclone model \citep{Bjerknes1919,Bjerknes1922} fronts were understood as zero-order discontinuity in the temperature field, which is until today the motivation for weather services to draw fronts as lines in weather maps. However, this view on fronts violates the basic continuity principle of fluid dynamics \citep[e.g.][]{Tao2014}, so it has been revisited by \citet{Bjerknes1937} and fronts are considered as narrow transition zones characterized by a strong horizontal temperature gradient. This property is often referred to as defining (or "primary") quantity for front detection \citep{Thomas2019}. In addition, dynamical processes occurring along fronts (see Introduction), lead to further characteristic features of fronts -- sometimes referred to as "secondary" quantities \citep{Thomas2019} -- such as cyclonic vorticity, strong vertical motion, enhanced stability and a positive potential vorticity (PV) anomaly \citep[e.g.][]{Hoskins1982}. \\
    Since all these properties are not unique features of fronts (e.g. a strong horizontal temperature gradient can occur along coastlines due to differential heating of land and sea), several of these quantities and their combinations have been utilized for front identification \citep[e.g.][]{Renard1965,Simmonds2011,Parfitt2017}, out of which the thermal front parameter (TFP, Sec. \ref{kap:intro-tfp}) \citep{Renard1965} and the F diagnostic (Sec. \ref{kap:intro-f}) \citep{Parfitt2017} are the most used. The TFP method is based on a primary quantity only, while the F diagnostic additionally takes a secondary quantity into account. However, both methods do not have a natural basic state, therefore we introduce the dynamic state index (DSI, Sec. \ref{chap:dsi}), which is an exact measure of the deviation from a general non-linear wind balance, for front detection. \\
	Generally, front detection methods are used to filter out frontal zones and front lines from gridded data. The procedure for this consists of three steps: (1) using one or more meteorological input variables, (2) apply a function to them, and (3) test for a threshold exceedance \citep{Hewson1998}. At the end, graphical filters can be applied, e.g. to delete disconnected pixels \citep[e.g.][]{Berry2011,Kern2019}. 

	\subsection{The case study: A rapid cyclogenesis, 14.02.2014} \label{kap:case-study}
\begin{figure}[H]
		\centering
		\includegraphics[width=14cm]{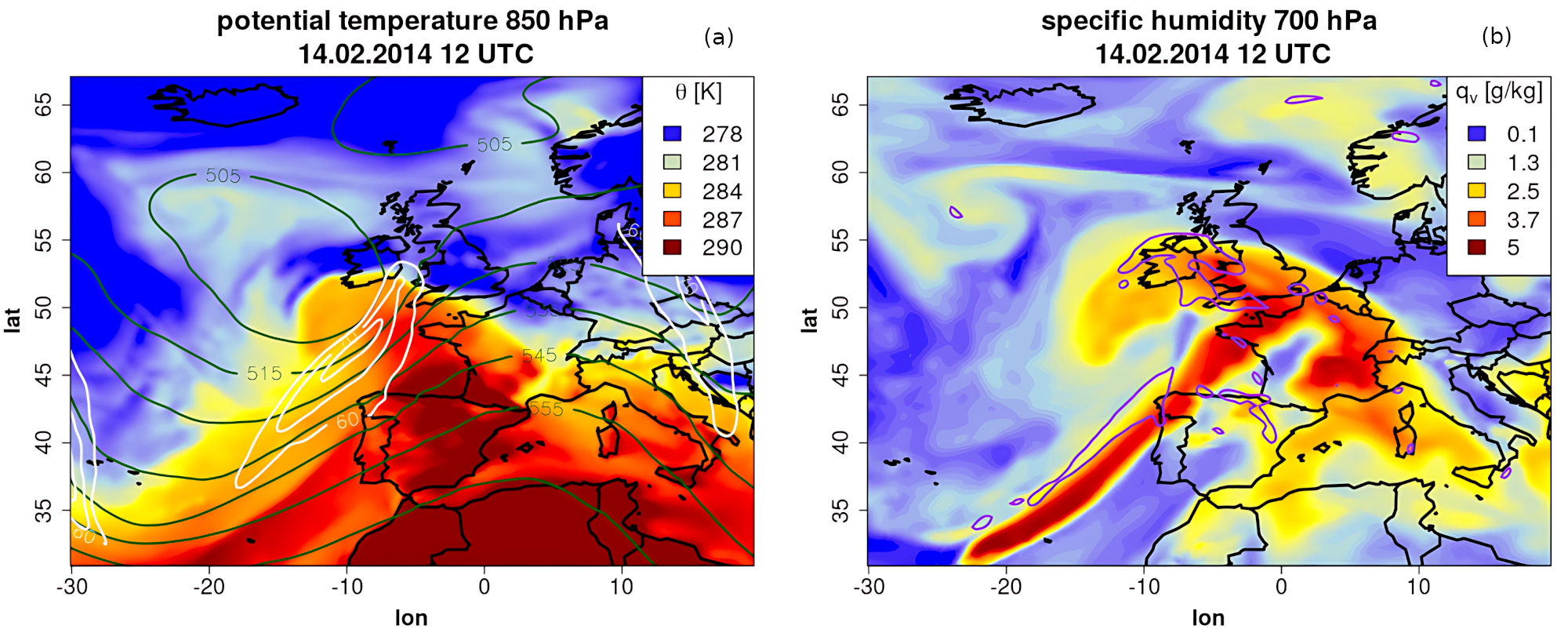}
		\caption{Weather situation 14.02.2014 12 UTC. (a) Potential temperature at 850 hPa with contour lines of geopotential height at 500 hPa (darkgreen, in gpdm) and of the horizontal wind at 300 hPa (white, in m/s). (b) Specific humidity at 700 hPa with contour line of -1 Pa/s updraft (purple).}
		\label{abb_synoptik}
	\end{figure}

    In a rapid cyclogenesis, diabatic processes play -- in addition to baroclinic instability -- a central role for the cyclone intensification \citep[e.g.][]{Wernli2024}. Thereby, mid-tropospheric diabatic heating contributes to a coupling of an upper-level PV anomaly, caused by a dry intrusion of stratospheric air into the troposphere, with a low-level PV anomaly, forming a "PV tower" \citep[e.g.][]{Hoskins1990,Rossa2000}. Here, we use a rapidly intensifying cyclone as a case study which has already been analyzed intensively with a focus on precipitation caused during its life cycle \citep{Bott2016}. \\
    Fig. \ref{abb_synoptik} shows a trough west of Ireland at 14.02.2014 12 UTC with a low pressure system on its south-eastern side, which is partly occluded and clearly shows a warm front (over France) and a cold front (west of Portugal). The jet stream flows over the cold front, so that it is cata-type in the northern part and ana-type in the southern part. Along the fronts the specific humidity is particularly high. Maxima of the vertical velocity can be seen in the area of the occlusion and at the back of the ana cold front. This cyclone was part of a series of rapid intensifying cyclones over the North Atlantic in February 2014 \citep{Volonte2018}.
 
	\subsection{Method 1: Thermic front parameter} \label{kap:intro-tfp}
\begin{figure}[H]
		\centering
		\includegraphics[width=7cm]{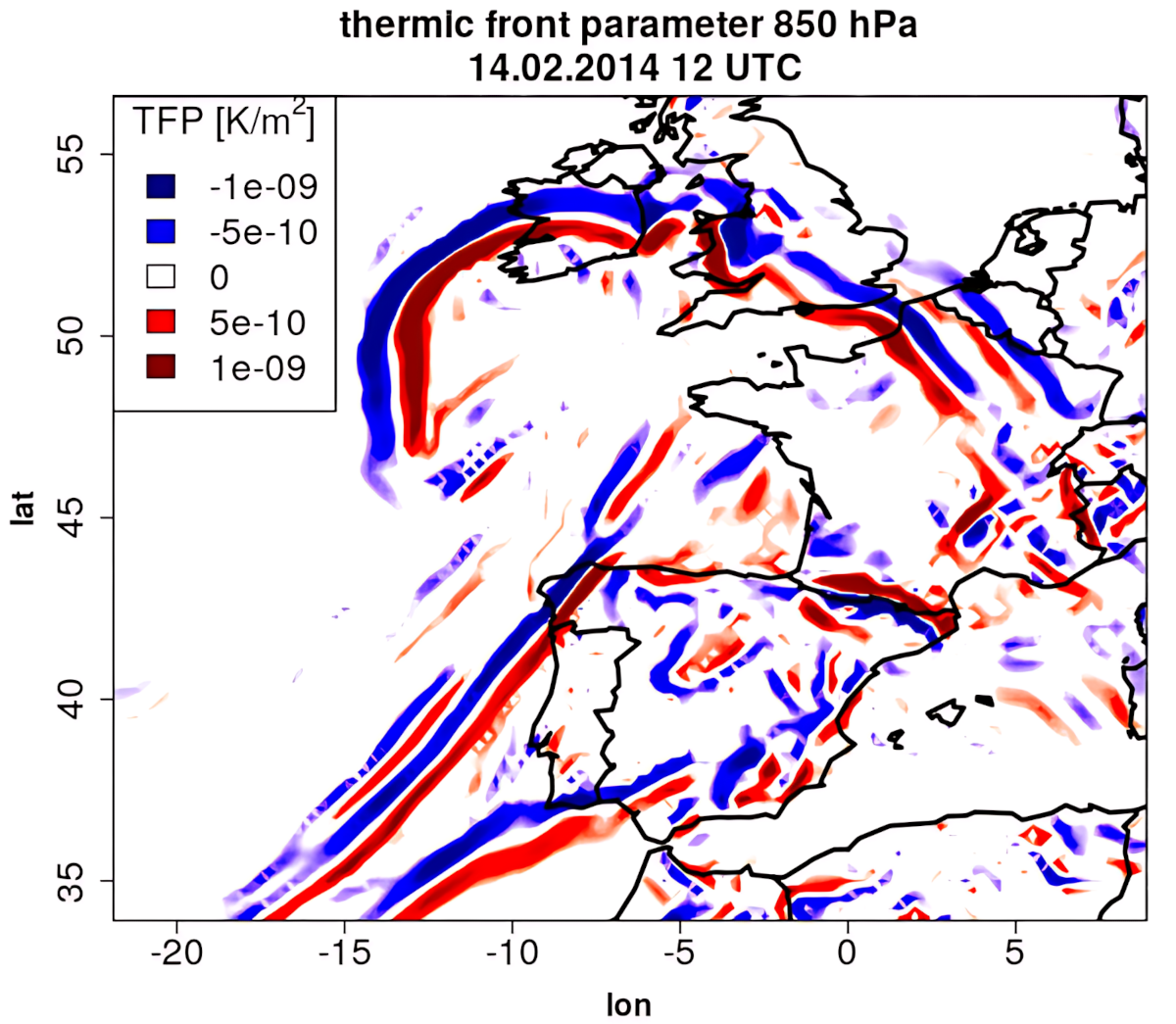}
		\caption{Thermic front parameter (TFP) at 850 hPa, 14.02.2014 12 UTC.}
		\label{abb_tfp}
	\end{figure}

    Fronts as air mass boundaries are characterized by high baroclinicity, i.e. a pronounced horizontal gradient $\nabla_h \tau$ in a temperature-based quantity $\tau$. 
    To formalize this approach to an automatic detection method, \citet{Renard1965} introduced the thermal front parameter (TFP) through
	\begin{equation}
		TFP({\tau}) := -\nabla_h \Vert\nabla_h \tau\Vert \cdot \dfrac{\nabla_h \tau}{\Vert\nabla_h \tau\Vert}.
	\end{equation}
	The first factor is given by the horizontal gradient of the baroclinicity and thus represents horizontal variations in baroclinicity. The second factor is the horizontal unit vector along the thermal gradient, and through the scalar multiplication of both terms only their parallel projection is considered, such that the TFP is a measure for the horizontal change of baroclinicity along the thermal gradient. For $\tau$ different thermal quantities can be considered, usually either potential temperature $\theta$ \citep[e.g.][]{Berry2011} or equivalent-potential temperature $\theta_e$ \citep[e.g.][]{Schemm2015}. \citet{Thomas2019} showed, that using $\tau=\theta_e$ for front detection leads to artificial fronts in the tropical regions due to high moisture content. To avoid this problem we use $\tau=\theta$ in accordance with e.g. \citet{Berry2011}.
	Fig. \ref{abb_tfp} shows the TFP at 850 hPa exemplary for the considered case study. The TFP has a dipole structure, with negative values at the front side of the warm front and the occluded front, and positive values at the front side of the cold front. The cata cold front is not visible in the TFP, so the cold front is detached from the frontal system, which is a characteristic feature of Shapiro-Keyser lows \citep{Shapiro1990}.

\begin{figure}[H]
		\centering
		\includegraphics[width=14cm]{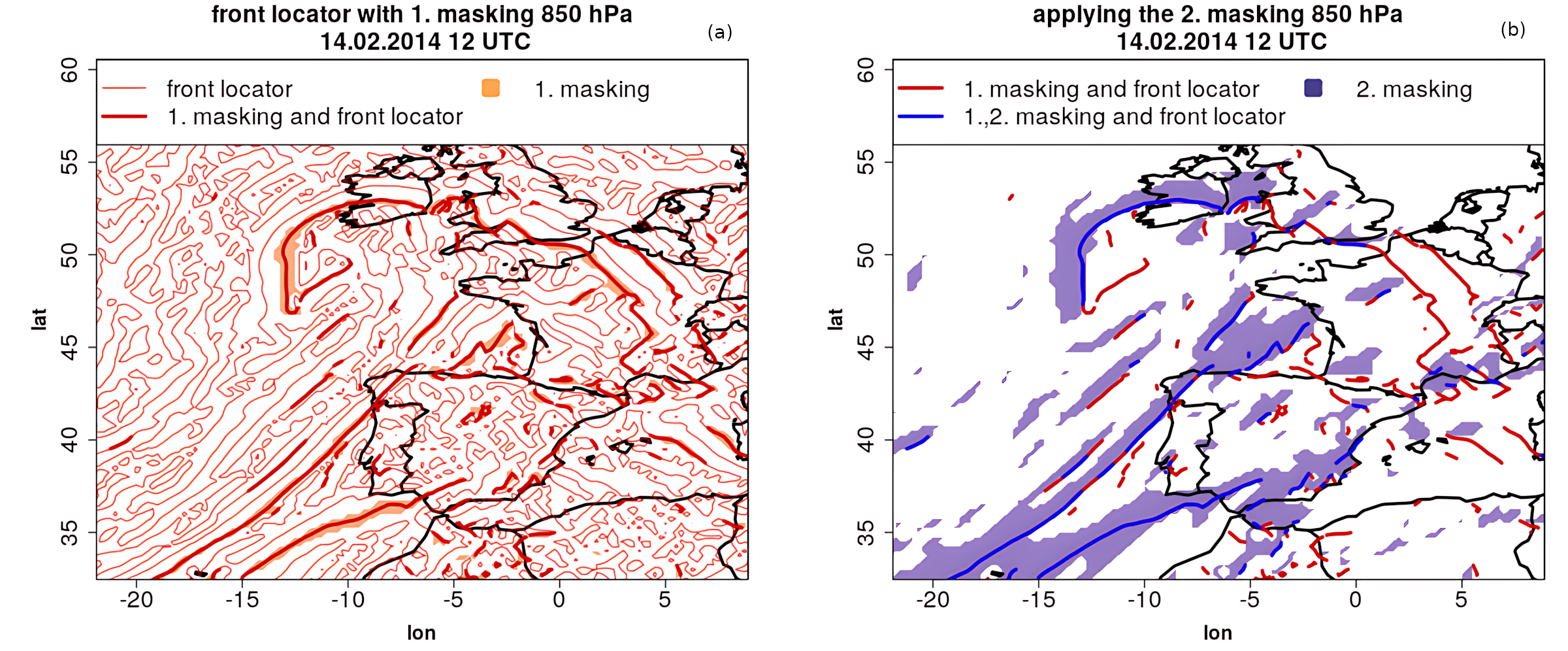}
		\caption{Front identification using the TFP method. (a) Front locator (thin red line), the first masking condition (orange shade) and the intersection of both (thick red line). (b) Results from the first masking step (thick red line), the second masking condition (blue shade) and the intersection of both (thick blue line).}
		\label{abb_frontid-tfp}
	\end{figure}

	Using the TFP concept, \citet{Hewson1998} developed an algorithm that allows the detection of fronts as front lines as well as frontal zones (algorithm 1):
    \begin{algorithm} \label{algo_tfp}
		\vspace{0.1cm}
		\textbf{Algorithm 1:} Front identification using TFP according to \citet{Hewson1998} 
		\hrule
		\vspace{0.1cm}
		\textbf{front lines}
		\vspace{-0.2cm}
		\begin{align}
		\textrm{front locator:} \hspace{1cm}&\frac{\partial(\Vert\nabla_h\Vert\nabla_h \tau \Vert \Vert)}{\partial \hat{s}} = 0 \;\textrm{ with }\; \hat{s} = \frac{\nabla_h \Vert\nabla_h \tau\Vert}{\Vert\nabla_h\Vert\nabla_h \tau \Vert \Vert} \label{gl_lokator1}\\
		& \nabla TFP(\tau) \cdot \frac{\nabla_h \tau}{\Vert \nabla_h \tau \Vert} = 0 \label{gl_lokator2} \\
		\textrm{masking variable 1:} \hspace{1cm}&TFP({\tau}) > K_1 \label{gl_mask1} \\
		\textrm{masking variable 2:} \hspace{1cm}& S({\tau}) := \Vert \nabla_h \tau \Vert > K_2 \label{gl_mask2}\\
		\vdots \hspace{3cm}& \notag\\
		\textrm{masking variable n}  \hspace{1cm} \notag&
		\end{align}
		\vspace{-0.6cm}
		\hrule
		\vspace{0.1cm}
		\textbf{frontal zones} \\[-0.6cm]
		\begin{align}
			TFP({\tau}) > K_1 \label{gl_tfp-zone}
		\end{align}
	\end{algorithm}
	First, the location of the front line is determined. In the frontal zone, the norm of the gradient of the thermal quantity $\tau$ is the largest, so that the first partial derivative has a maximum there. The second spatial derivative has a minimum at the front of the frontal zone (which is called front line) and the third derivative a root, which must be determined (see \citet{Kern2019} their Fig. 2 and 3). In order to additionally take into account the curvature of the front, the divergence along the unit vector $\hat{s}$ is considered resulting in Eq. \ref{gl_lokator1}, which was simplified by \citet{HuberPock1981} to Eq. \ref{gl_lokator2}. To filter out the front, masking conditions are applied. The first masking condition (Eq. \ref{gl_mask1}) states that the TFP must have a minimum value of $K_1$. The value of $K_1$ depends on the used thermal quantity, the pressure level and the grid size. Here we follow \citet{Kern2019} and use $K_1 = 0.3 \cdot 10^{-10}$ K/m for 850 hPa level, which is similar to \citet{Hewson1998} and \citet{Parfitt2017} ($K_1 = 0.33 \cdot 10^{-10}$ K/m for 900 hPa) but different from \citet{Catto2013} ($K_1 = -8 \cdot 10^{-12}$ K/m for the wet-bulb potential temperature at 850 hPa). Fig. \ref{abb_frontid-tfp}a shows the front locator (thin red lines), the first masking condition (orange shading) and the intersection of both resulting in the (preliminary) front lines (thick red lines). The occlusion front, warm front, ana cold front and also a part of the cata cold front displaced to the ana cold front are detected. To further filter the front lines, \citet{Hewson1998} used the strength of the front $S(\tau)$ given by Eq. \ref{gl_mask2}. If the second masking condition is applied (Fig. \ref{abb_frontid-tfp}b), then only the part of the (preliminary) front lines from the first masking step are kept which also show a minimum strength. Due to the lower baroclinicity, the warm front and a part of the occlusion front are now not captured, even when using a comparatively small threshold of $K_2 = 1.35 \cdot10^{-5}$ K/m \citep{Hewson1998,Parfitt2017,Kern2019}. Since this is not intentional, we use only the first masking step in this study. This also allows filtering for frontal zones directly (Eq. \ref{gl_tfp-zone}). Consequently, warm fronts are located at the end of the warm air advection and cold fronts at the beginning of the cold air advection. 

	\subsection{Method 2: F diagnostic} \label{kap:intro-f}
\begin{figure}[H]
	\centering
	\includegraphics[width=14cm]{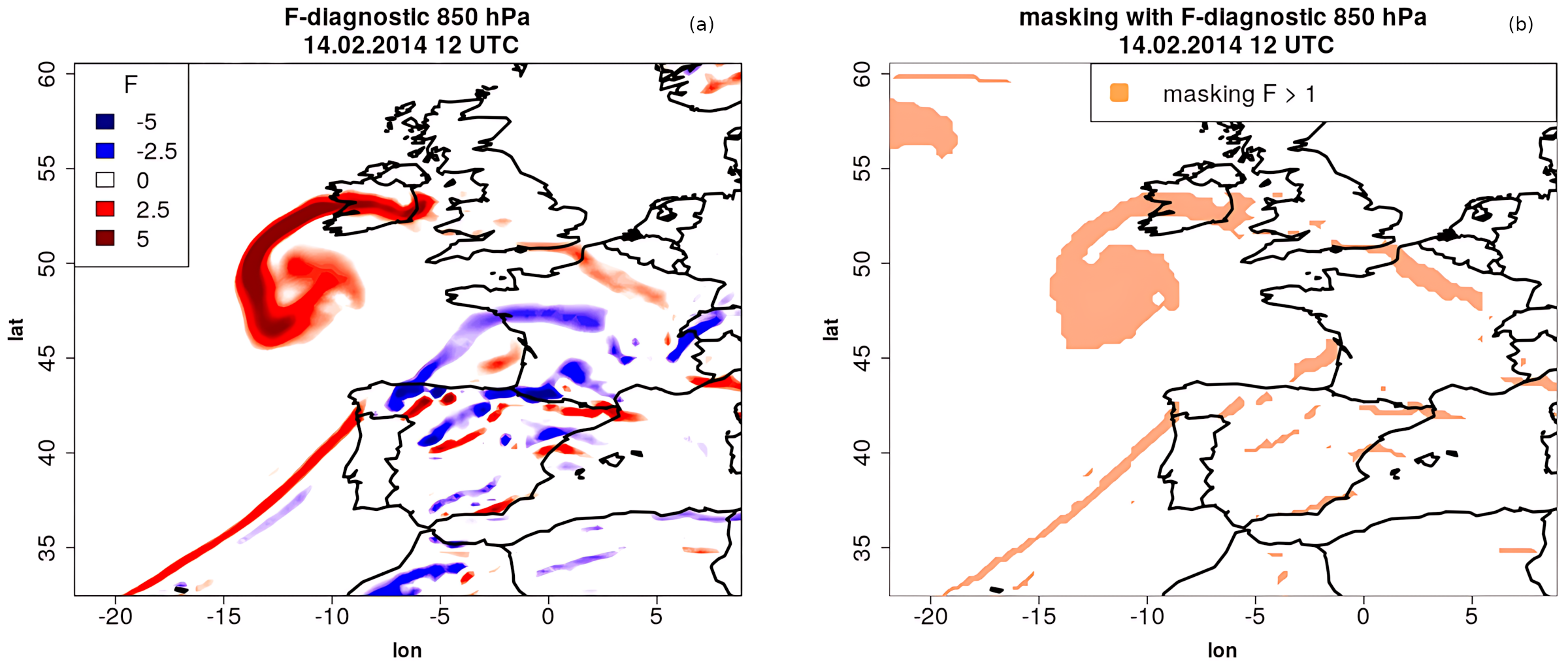}
	\caption{Front identification using the F diagnostic. (a) F diagnostic at 850 hPa. (b) Masking condition $F>1$.}
	\label{abb_frontid-f}
\end{figure}

	\citet{Parfitt2017} use the fact that fronts are regions of strong temperature gradients \textit{and} maximal vorticity to develop a new front identification method (algorithm 2), which only allows the detection of frontal zones (and not lines):
\begin{algorithm}
	\vspace{0.1cm}
	\textbf{Algorithm 2:} Front identification using F diagnostic according \\to \citet{Parfitt2017}
	\vspace{0.1cm}
	\hrule
	\begin{align}
	\textrm{defining quantity:} \quad F &:= \frac{\zeta}{f}\frac{\Vert\nabla_h T\Vert}{\Vert\nabla_h T\Vert_0} \\
	\textrm{masking:} \quad F &\begin{cases}
	> 1, &\textrm{for low-level fronts}\\
	> 2, &\textrm{for upper-level fronts (from 600 hPa)} \label{gl_mask-parfitt}
	\end{cases}
	\end{align}
	\vspace{-0.2cm}
\end{algorithm}
F is a empirically determined and dimensionless quantity containing the relative vorticity $\zeta$ (on isobaric surfaces), the Coriolis parameter $f$, the temperature $T$ and a empirical determined constant $\Vert \nabla T \Vert_0 = 0.45$ K/(100 km). The threshold values used for masking (Eq. \ref{gl_mask-parfitt}) were also determined empirically based on ERA-Interim \citep{Parfitt2017}. \\
We point out that F contains the Rossby number
\begin{align} \label{gl_ro-zahl}
Ro := \frac{\zeta}{f} \begin{cases}
\ll 1 \;&\rightarrow \textrm{synoptic scale} \\
\approx 1 \;&\rightarrow \textrm{mesoscale} \\
\gg 1 \;&\rightarrow \textrm{convective scale},	
\end{cases}
\end{align}
which describes the ratio of inertia to Coriolis force and can be interpreted as a measure for the spatial scale. Consequently, by using the F diagnostic for front detection, fronts are perceived on the basis of their spatial scale weighted with the horizontal temperature gradient. The F diagnostic (Fig. \ref{abb_frontid-f}a) shows high values in the low center and along the occlusion front due to high relative vorticity caused by the dry intrusion, which generates potential instability and thus can initiate convection \citep{Browning1997}. Positive F values also occur along the ana cold front and at a small part of the warm front, while negative F values are inevitable connected to negative relative vorticity. If now the masking condition $F>1$ is applied, the frontal zones shown in Fig. \ref{abb_frontid-f}b result. Thereby, the low center is erroneously detected as a front and the warm and cold fronts are detected only to a small part detached from the low center. Since baroclinicity and relative vorticity are not independent of each other but are positively correlated \citep{Hoskins1982}, only particularly strong fronts are detected with the F diagnostic, while weaker fronts remain unrecognized.

	\subsection{Method 3: Dynamic State Index (DSI)} \label{chap:dsi}

	In this section the concept of the dynamic state index (DSI) is introduced based on the derivation of a general non-linear solution of the primitive equations, called steady wind. Using the case study from before the DSI structure of the fronts is showcased, which is then used to develop a new algorithm for front identification by applying the DSI as masking variable.
	
	\subsubsection{The concept of the dynamic state index (DSI)}
\begin{figure}[H]
		\includegraphics[width=11cm]{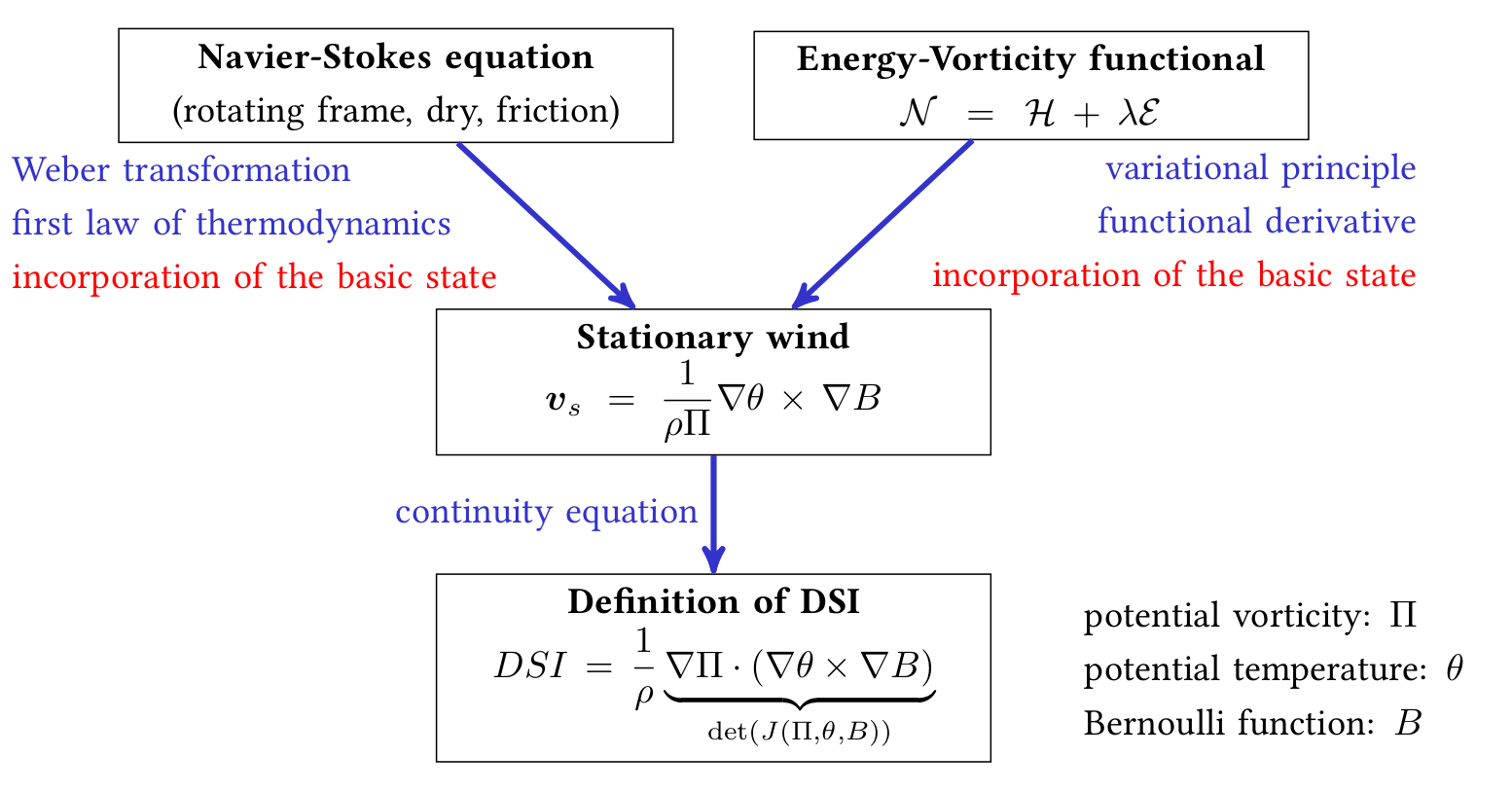}
		\centering
		\caption{Scheme that shows the different ways to derive the DSI. The two left arrows (i.e., derivation of steady wind from Navier-Stokes equation, and derivation of DSI from inserting the steady wind in the continuity equation) were detailed in the text.}
		\label{abb_dsi-derivation}
	\end{figure}

	In order to investigate flows systematically, they are often divided into a basic state and a deviation from it. This basic state can, for example, be determined statistically with the mean value. The deviation is then given in first order by the standard deviation. Such a statistical approach has the disadvantage that the mean value does not necessarily represent a solution of the basic equation system and is therefore not balanced. For this reason, a dynamic basic state is desirable which itself represents a solution to the primitive equations. Here, we consider an adiabatic and stationary basic state leading to the steady wind solution of the primitive equations \citep{Schaer1993} and the deviations from it are given by the DSI \citep{Nevir2004, Weber2008}. The advantage of this "dynamical" approach is that the basic state is balanced, however, it  has to be derived first. The derivation procedure is summarized in Fig. \ref{abb_dsi-derivation} and detailed in the following. 
 
    \subsubsection{The steady wind representation and its interpretation}
	The derivation of the steady wind starts directly from the primitive equations (all steps are detailed in appendix \ref{appendix:steady-wind}). The first law of thermodynamics is incorporated into the Navier-Stokes equation leading to an alternative version, that combines momentum and energy conservation in one equation. Then the basic state (adiabatic, steady, inviscid) is incorporated, which means that diabatic, instationary and friction terms are omitted. This yields an expression for the convective flux of the PV \citep{Schaer1993} $\rho \Pi \boldsymbol{v} = \nabla \theta \times \nabla B $, which can be rearranged to
	\begin{equation}\label{gl_statwind}
	\boldsymbol{v}_{s} = \frac{1}{\rho \Pi} \nabla \theta \times \nabla B, 
	\end{equation}
	what is called the steady wind representation by \citet{Schaer1993} and \citet{Nevir2004}. It contains four constituting quantities: the air density $\rho$, the potential temperature $\theta$, the Bernoulli function $B:= \Phi + 0.5\, \boldsymbol{v}^2 +h$, which represents the most general stream function as the sum of potential energy $\Phi$, kinetic energy $0.5\,\boldsymbol{v}^2$ (based on the wind vector $\boldsymbol{v}=(u,v,w)^T$) and enthalpy $h$. The potential vorticity $\Pi := \rho^{-1} \boldsymbol{\xi_a} \cdot \nabla \theta$ (utilizing the absolute vorticity $\boldsymbol{\xi_a}$) describes the circulation on isentropic surfaces and is a central quantity in atmospheric vorticity dynamics \citep{Ertel1942,Hoskins1985}. \\
    Under adiabatic and stationary conditions, $\boldsymbol{v}_s$ does not intersect the tubes spanned by $\nabla B$ and $\nabla \Pi$ and therefore, does neither constitute to the PV tendency nor to energy conversions, which is why \citet{Gassmann2014, Gassmann2019} refers to it as "inactive" wind. The real wind can be decomposed by $\boldsymbol{v}=\boldsymbol{v}_s+\boldsymbol{v}_a$ into steady (inactive) wind and the deviation from it, the so-called active wind $\boldsymbol{v}_a$ \citep{Gassmann2014}. Compared to the geostrophic wind decomposition, the active wind can be seen as analogy to the ageostrophic wind. The vertical component of the steady wind $w_s$ describes isentropic upglide, while the deviation from it $w-w_s$ behaves similarly to the isentropic displacement wind \citep{Hoskins2003,Gassmann2019}, which affects baroclinicity by tilting of isentropic surfaces \citep{Papritz2015}. \\
	A comparison of the steady wind $\boldsymbol{v}_s$ with other wind representations reveals a hierarchy in terms of the contributing energetic and vortical quantities (Tab. \ref{tab_wind-representations}). The geostrophic wind $\boldsymbol{v}_{g,h}$ contains as energetic quantity the potential energy $\Phi$, which acts as stream function, and as vortical quantity the Coriolis parameter $f$, which represents the external rotation only. The pseudo-geostrophic wind $\boldsymbol{v}_{pg,h}$ (derived under the assumption $\partial \boldsymbol{v}_h/\partial t=0$) contains the sum of potential and kinetic energy as energetic quantity and the absolute vorticity $\zeta + f$, i.e. external and internal rotation, as vortical quantity \citep{Lange2002}. The kinetic energy (following from considering the advection term) introduces non-linearity, while the geostrophic wind solution is linear.
    The steady wind $\boldsymbol{v}_s$ additionally includes thermodynamic information, such that the energetic quantity is the Bernoulli function and the vortical is the "PV substance" $\rho \Pi$. $\boldsymbol{v}_s$ is thus the most general wind representation, which contains all external and internal energetic and vortical aspects \citep{Schaer1993,Lange2002,Nevir2004}. \\
    The fact that all the wind representations contain an energetic and a vortical quantity shows that the two are equally important conserved quantities and represent constraints to the flow, which reduce the phase space dynamics. Based on this observation, a generalized theory can be derived -- the energy-vorticity theory (EVT) -- which unifies the separate ways of thinking of a pure energy-based theory (Hamilton mechanics) and a pure vorticity theory (PV thinking) into a common theory based on Nambu mechanics \citep{Nevir1993,Nevir2009}. In the EVT framework the steady wind can also be derived by minimizing the energy-vorticity functional containing total energy and potential enstrophy \citep[e.g.][]{Weber2008}.
    
   \begin{table} \label{tab_wind-representations}
       \centering
       \begin{tabular}{l|ccc}
        &\textbf{geostrophic wind} & \textbf{pseudo-geostrophic wind} & \textbf{steady wind} \\
        \hline
            formula& 
            $\boldsymbol{v}_{g,h} = \dfrac{1}{\textcolor{darkblue}{f}} \boldsymbol{k}\times \nabla_h \textcolor{darkgreen}{\Phi}$ & 
            $\boldsymbol{v}_{pg,h} = \dfrac{1}{\textcolor{darkblue}{\zeta+f}} \boldsymbol{k} \times \nabla_h \left( \textcolor{darkgreen}{\dfrac{1}{2} \boldsymbol{v}_h^2 + \Phi} \right)$ & 
            $\boldsymbol{v}_{s} = \dfrac{1}{\textcolor{darkblue}{\rho \Pi}} \nabla \theta \times \nabla \textcolor{darkgreen}{B}$\\
            \textcolor{darkblue}{vortical quantity}& \textcolor{darkblue}{$f$} & \textcolor{darkblue}{$\zeta+f$} & \textcolor{darkblue}{$\rho \Pi$}\\
            \textcolor{darkgreen}{energetic quantity} & \textcolor{darkgreen}{$\Phi$} & \textcolor{darkgreen}{$\dfrac{1}{2} \boldsymbol{v}_h^2 + \Phi$} & \textcolor{darkgreen}{B}\\
            dimensions &2&2&3\\
       \end{tabular}
       \caption{Different common wind approximations (geostrophic wind, pseudo-geostrophic wind and steady wind) and the contributing energetic and vortical quantities compared. The complexity increases from left to right with the steady wind as most general wind representation.}
       \label{tab:my_label}
   \end{table}

    \subsubsection{The Dynamic State Index (DSI)}
	For the derivation of the DSI (summarized in Fig. \ref{abb_dsi-derivation}), we insert $\boldsymbol{v}_s$ in the continuity equation
	$
	\nabla \cdot (\rho \boldsymbol{v}_s) = 0,
	$
	yielding
	$
	\nabla \Pi \cdot \left(  \nabla \theta \times \nabla B   \right) = 0.
	$
	This motivates the definition of the dynamic state index DSI \citep{Nevir2004} by
	\begin{align}\label{gl_dsi-def}
	DSI := \frac{1}{\rho} (\nabla \theta \times \nabla B) \cdot \nabla \Pi &= \frac{1}{\rho} \det \left(\frac{\partial (\theta, B, \Pi)}{\partial (x,y,z)} \right) \hspace{1cm} \textrm{(determinant form).}
	\end{align}
	In the second equality the triple product is reformulated in terms of the determinant of the Jacobian. The DSI combines the Lagrangian conserved quantities $\theta$ and $\Pi$ with the most general stream function $B$ and measures their phase space volume. In this sense, it can be seen as a generalization of the PV-$\theta$ view on atmospheric circulation \citep{Hoskins1991}.\\
    By applying the product rule the DSI can equivalently be written as 
	\begin{align}
		DSI &=-\frac{\Pi^2}{\rho} \nabla \cdot (\rho \boldsymbol{v}) \hspace{1cm} \textrm{(divergence form)} \\
		&=\boldsymbol{v} \cdot \nabla \left( \frac{\Pi^2}{2} \right)	 \hspace{1.1cm} \textrm{(advection form).} \label{gl_dsi-advektion}
	\end{align}
	The divergence form describes the continuity of the stationary wind and the advection form the advection of the squared PV, which can be considered as mass specific potential enstrophy. In the basic state with $\boldsymbol{v} = \boldsymbol{v}_s$, the divergence $\nabla \cdot \boldsymbol{v}_s$ and the advection of potential enstrophy vanish, such that $DSI = 0$. Furthermore, the advection of $\nabla \theta, \nabla B$ and $\nabla \Pi$ disappear in the basic state. At the same time, PV is a function of $B$ in the basic state, leading to a coupling of energetic and vortical quantity and thus a reduced non-linearity. This coupling between PV and B is a prerequisite in the PV thinking, but a dynamical evolution of the system is only possible if PV and B are independent from each other, that is, exactly when $DSI\ne 0$ \citep{Lange2002}. 
    Overall, the DSI values can precisely be interpreted by
	\begin{equation}\label{gl_dsi-interp}
	DSI \begin{cases}
	= 0, \;\;\; \textrm{stationary, adiabatic, reversible and frictionless flow (i.e. basic state)}\\
	\ne 0, \;\;\; \textrm{non-steady, diabatic or viscous flow}.
	\end{cases}
	\end{equation}

	\subsubsection{Using the DSI for front identification}
\begin{figure}[H]
		\centering
		\includegraphics[width=14cm]{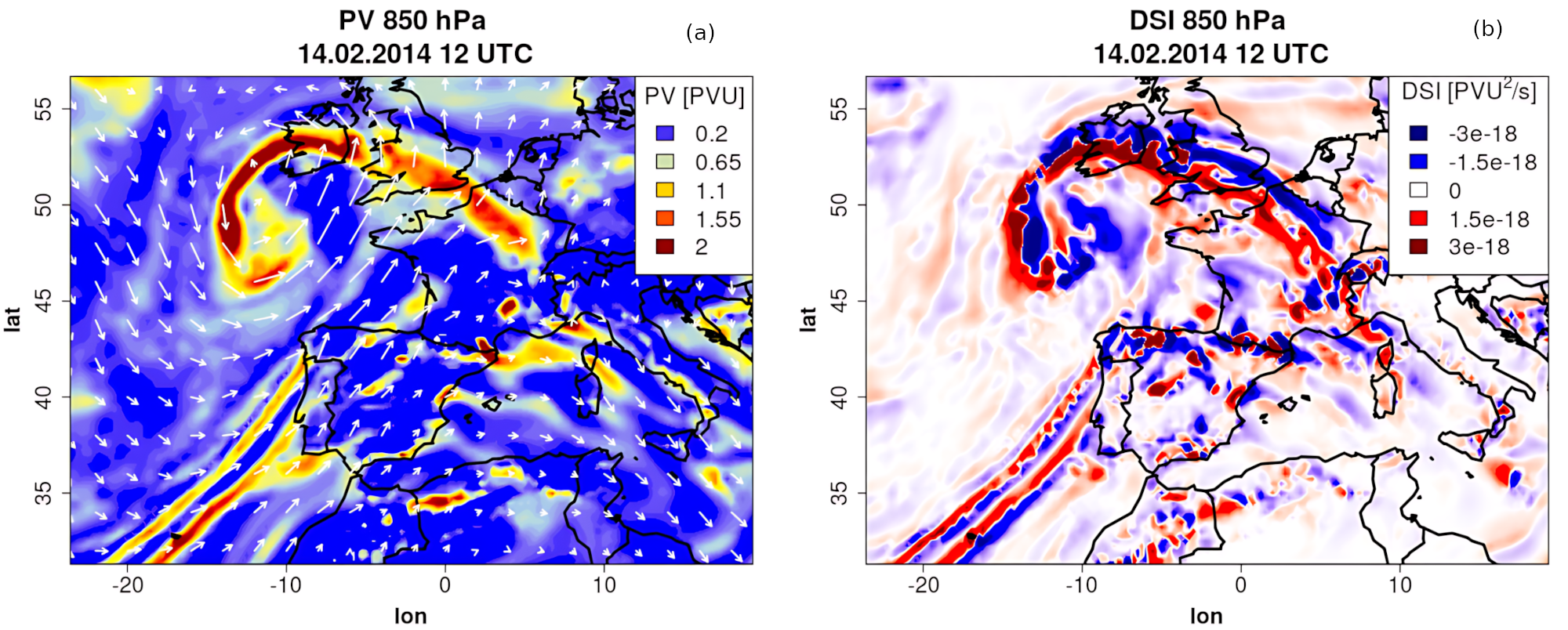}
		\caption{(a) Potential vorticity (coloured) and horizontal wind (white arrows) both at 850 hPa. (b) DSI at 850 hPa.}
		\label{abb_dsi-pv}
	\end{figure}

	Fig. \ref{abb_dsi-pv} shows the PV together with the horizontal wind and the DSI for the considered case study. The PV is maximal in the frontal regions due to large cyclonic vorticity, a large horizontal temperature gradient and strong stability. The PV anomalies along fronts arise from diabatic heating, e.g., due to condensation in the warm conveyor belt, evaporation of precipitation in the dry intrusion, convection, turbulent mixing and longwave radiative effects, as in detail studied based on PV tendencies by \citet{Chagnon2013} and \citet{Attinger2021}.
    The DSI shows a dipole structure along the fronts with negative values downstream and positive values upstream, reflecting an oscillation around the basic state, i.e. the energy-vorticity equilibrium. To explain this structure, the advection form of the DSI (Eq. \ref{gl_dsi-advektion}) is considered.
	Based on the wind decomposition in steady (inactive) and active wind through $\boldsymbol{v} = \boldsymbol{v}_s + \boldsymbol{v}_a$, \citet{Gassmann2014} showed that in the horizontal (index $h$) the stationary wind dominates the active wind (i.e., $\boldsymbol{v}_{s,h} \gg \boldsymbol{v}_{a,h}$) and thus can be used as approximation for the real wind (i.e., $\boldsymbol{v}_{h} \approx \boldsymbol{v}_{s,h}$). Consequently, a zonal wind crossing a cyclonic PV anomaly leads to positive values upstream (due to $\nabla \Pi^2>0$) and negative values downstream (due to $\nabla \Pi^2<0$). Thus, frontal zones are characterized by large $|DSI|$ values. From inserting the wind decomposition into the continuity equation (\citet{Gassmann2014}, therein Eq. 2.24), it can further be concluded, that positive (normalized) DSI values depict a source for the active wind, while negative values depict a sink for the active wind \citep{Gassmann2014,Gassmann2019}. The vertical active wind (and the in shape similar isentropic displacement wind) pushes the isentropic surfaces down ahead of a front, leading to adiabatic warming (in the warm sector), while at the backside of the front the isentropes are pushed up, associated with adiabatic cooling (in the cold sector) \citep{Hoskins2003}. The horizontal active wind generates kinetic energy, when it points towards low pressure \citep{Gassmann2019} and can thus be seen as analogy to the ageostrophic wind, which contains the isallobaric wind (that "blows" perpendicular to lines of the same pressure fall tendency). From the simplification of viewing fronts as discontinuity surfaces it can be deduced, that fronts always move to the area of the greatest pressure fall tendency, which corresponds to the front-perpendicular component of the isallobaric wind \citep{Zdunkowski2003}.  
    In this way, the DSI can be used as a diagnostic indicating dynamically active regions, in analogy to the divergence of the $\boldsymbol{Q}$ vector, which -- derived from the quasi-geostrophic 'omega'-equation -- indicates vertical motions and is thus used as an indicator of frontogenesis \citep{Hoskins1978,Steinacker1992}.
    In addition to these theoretical considerations, data-based studies can help to better understand the behaviour of the DSI. \citet{Mueller2018} showed that the structure of frontal rain bands is reflected in the DSI, and based on time series analysis, \citet{Claussnitzer2009} and \citet{Claussnitzer2010} demonstrated that $|DSI|$ shows a high correlation with precipitation sums. Moreover, \citet{vonLindheim2021} showed that the DSI is associated with persistent and coherent structures in atmospheric flows, like fronts. All these properties make the DSI a suitable tool to identify fronts as deviation from the adiabatic and steady basic state. For this purpose, we have developed the following algorithm (algorithm 3):
\begin{algorithm}\label{algo_dsi}
	\vspace{0.1cm}
	\textbf{Algorithm 3:} Front identification using DSI (''DSI method'')
	\vspace{0.1cm}
	\hrule
	\vspace{0.1cm}
	\textbf{front lines} \\[-0.9cm]
	\begin{align}
	\textrm{front locator: } \quad & \Vert\nabla_h DSI\Vert \cdot \frac{\nabla_h \Pi}{\Vert \nabla_{h} \Pi\Vert }= 0 \label{gl_dsi-lokator} \\
	\textrm{masking: } \quad & DSI<q_{p_1}(DSI) 
    \label{gl_mask-dsi}
	\end{align}
	\vspace{-0.5cm}
	\hrule
	\vspace{0.1cm}
	\textrm{\textbf{frontal zones}}
	\begin{align}
    \textrm{masking: } \quad & |DSI|> q_{p_2}(DSI) 
    \label{gl_mask_dsi-zone}
	\end{align}
\end{algorithm}

\begin{figure}[H]
		\centering
		\includegraphics[width=7cm]{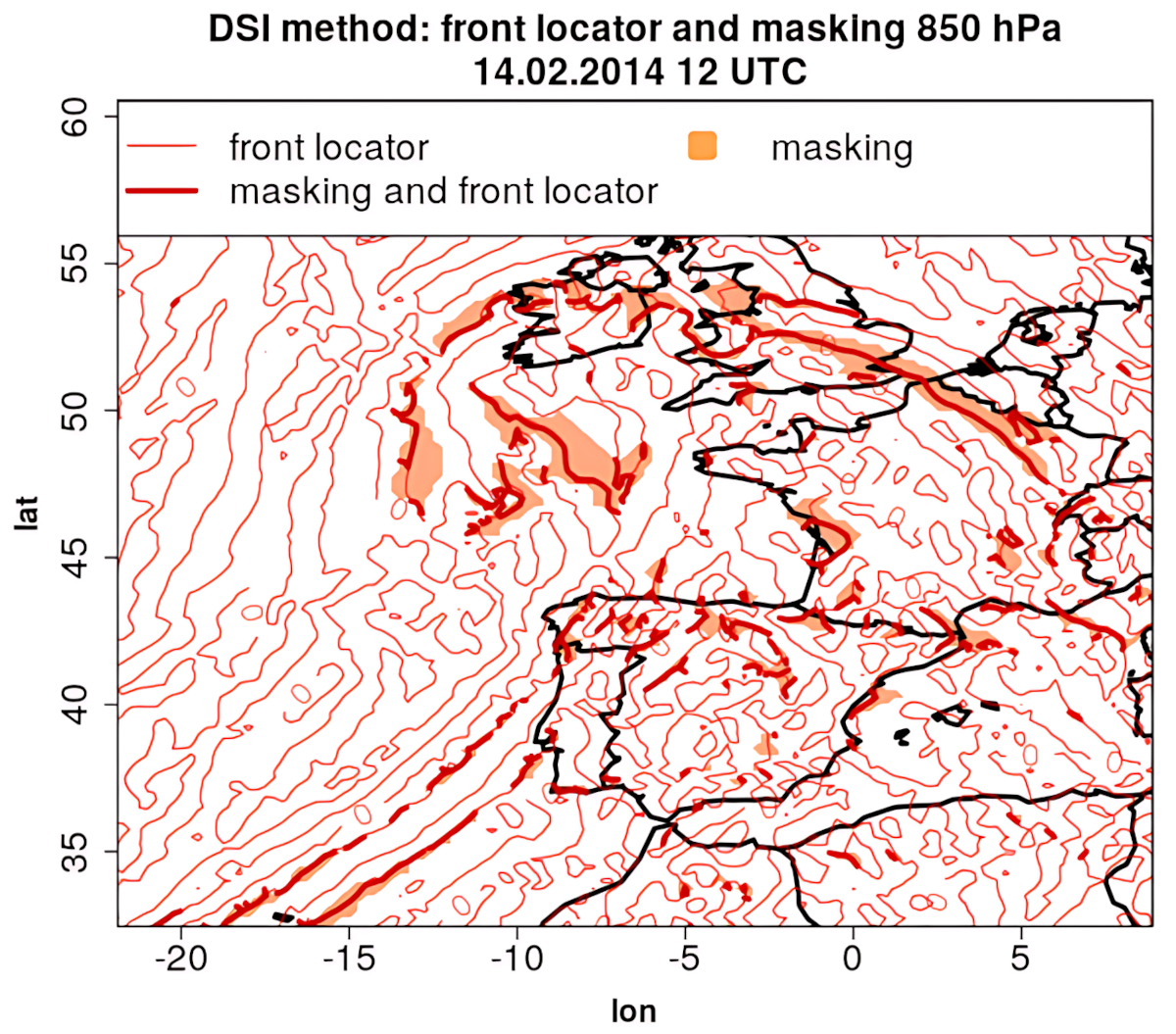}
		\caption{DSI method:  Front locator (thin red lines), masking condition (orange shading) and the intersection of both (thick red lines) at 850 hPa, 14.02.2014 12 UTC.}
		\label{abb_dsi-method}
	\end{figure}

	For the front locator, we consider the gradient of the DSI projected onto the unit vector of the PV. This is inspired by the TFP method (Eq. \ref{gl_lokator2}), but in contrast to the unit vector of the thermic quantity we use the unit vector of the PV, which always points in the direction of the frontal zone and consequently allows us to outline the whole frontal zone (front and back side). In order to filter out just the front line, the DSI must have particularly negative values, which can be measured by using a small percentile $q_{p1}$ (masking Eq. \ref{gl_mask-dsi}).  As Fig. \ref{abb_dsi-method} shows, the occlusion front, the warm front and the ana cold front are detected in the case study. Compared to the TFP method, the cold front line is more interrupted, and in contrast to the F diagnostic, the entire warm front is detected. \\
	Frontal zones are detected by particularly high values of the magnitude of the DSI (Eq. \ref{gl_mask_dsi-zone}), by using a large percentile $q_{p2}$. This is different from the TFP method, where only the positive values (and not the absolute values) are used to capture frontal zones. We determine the threshold values using an area over the North Atlantic (see Fig. \ref{abb_na-clim}) that is undisturbed by orography. The percentiles are calculated for every pressure level separately to account for the average increase in DSI with height.  Using a percentile allows to adjust the sensitivity of our method and to adapt it to the model resolution, since higher resolved models tend to have higher DSI values \citep{Weijenborg2015}.


\section{Results} \label{chap:comparison}
In this section, we compare the three methods for front identification (TFP, F diagnostic and DSI) spatially and temporally. As the F diagnostic only allows the detection of frontal zones, we consider them only and refer to them as ''fronts'' for the sake of simplicity. In the end, we examine the essential characteristics of the respective detected fronts. For this purpose, we consider mid-level fronts at 600 hPa. Using 850 hPa (as shown before) is unfavorable not only due to strong orography influence but also due to the influence of the boundary layer, which itself is strongly influenced by the land-sea contrast and the sea surface temperature \citep[e.g.][]{Parfitt2016}.

	\subsection{Climatology}
\begin{figure}
		\centering
		\includegraphics[width=13cm]{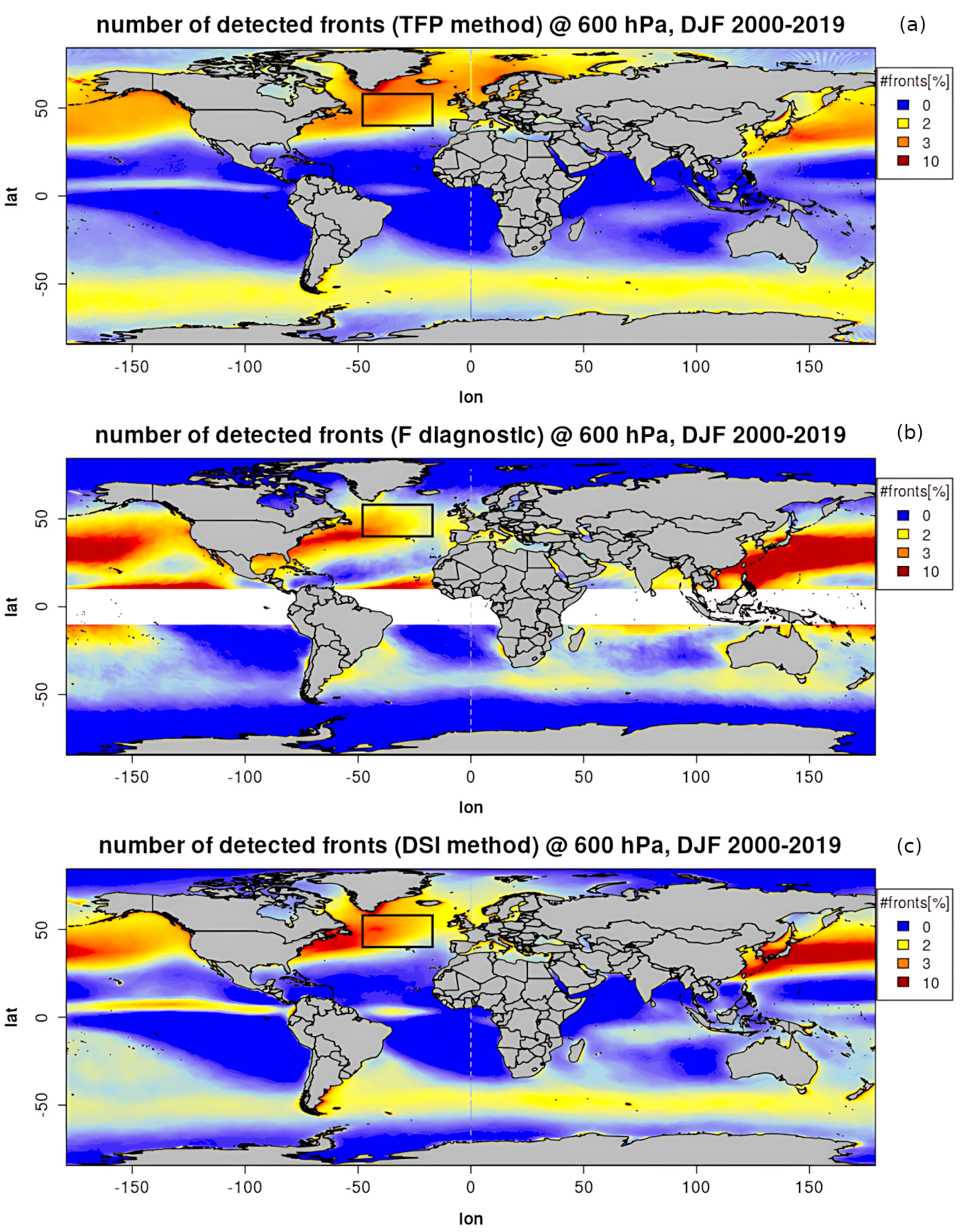}
		\caption{Front Climatology: Number of detected fronts (measured as occurrence probability) for twenty north winters (DJF 2000-2019) using (a) TFP method, (b) F diagnostic and (c) DSI method with 95 \% percentile.}
		\label{abb_na-clim}
	\end{figure}

	Fig. \ref{abb_na-clim} shows the occurrence probability of detected fronts (global north winter climatology, i.e. DJF 2000-2019, based on hourly ERA5 data) using the three previously described methods. 
	The TFP method detects the Northern Hemisphere Atlantic and Pacific storm tracks, which are partly washed out to the north. The Southern Hemisphere (summer) storm tracks are also recognizable in lower intensity. The F diagnostic (using the recommended threshold $F>2$ for mid-level fronts) shows strong signals along the Northern Hemisphere storm tracks, but only weak signals along the Southern Hemisphere storm tracks. The region around the equator is masked out due to the convergence of the Coriolis parameter, as recommended by \citet{Parfitt2017}. The DSI method detects clearly the Northern Hemisphere storm tracks with their typical characteristics -- northward tilt and eastward intensity decrease \citep{Hoskins2002}. The Southern Hemisphere storm tracks are detected by the DSI method as well, and in a stronger intensity than with the F diagnostic. Compared to the TFP method, the Southern Hemisphere storm tracks detected by the DSI method are more pronounced over the Atlantic and Indian Ocean, which is consistent with the cyclogenesis density found by \cite{Hoskins2005}. All methods show also signals along the inner tropical convergence zone.

	\subsection{Annual cycle}
\begin{figure}[H]
		\centering
		\includegraphics[width=8cm]{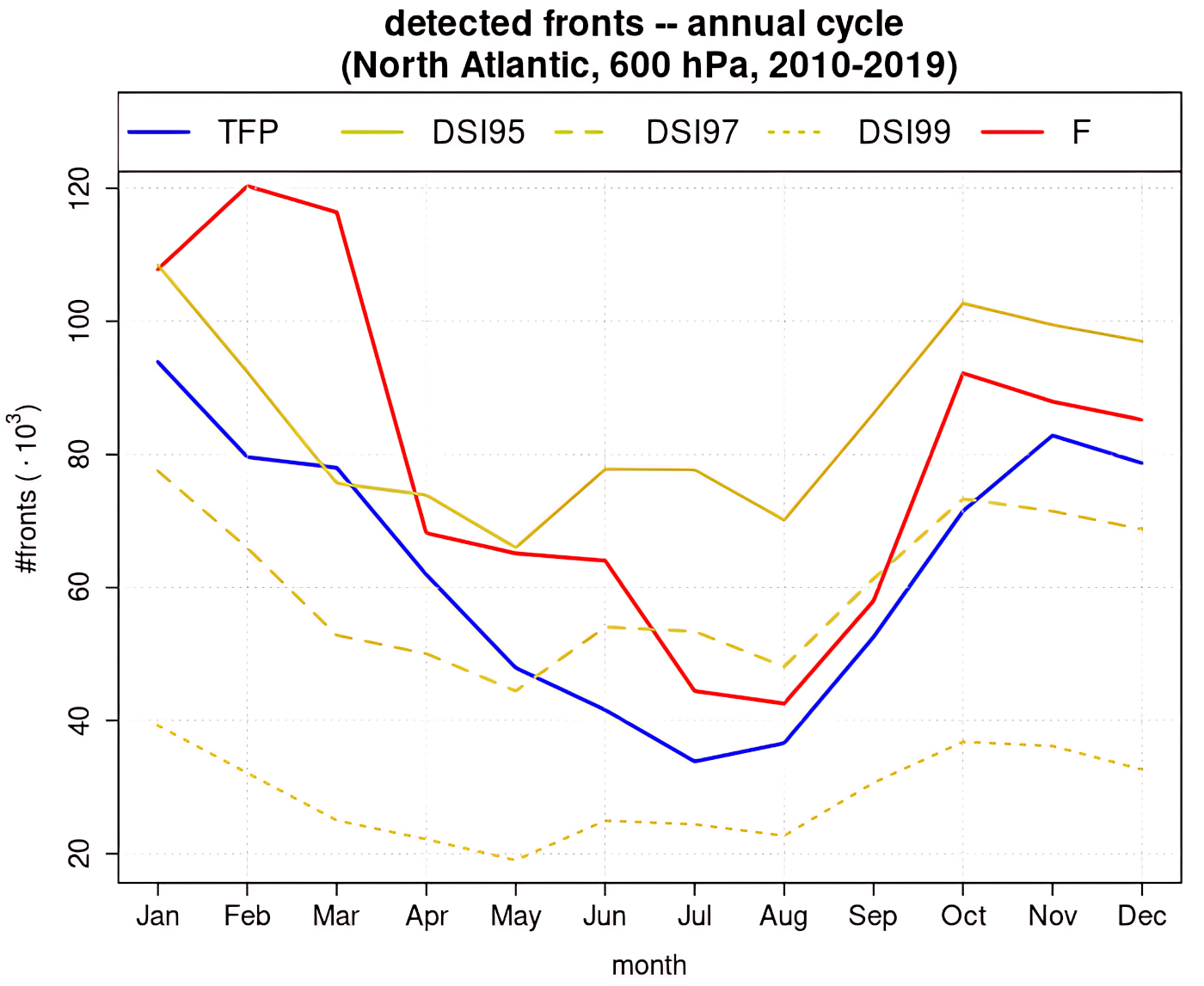}
		\caption{Annual cycle of detected fronts: Average number of detected fronts per month using TFP method, F diagnostic and DSI method (percentiles 95 \%, 97 \% and 99 \%, respectively) at 600 hPa over the North Atlantic between 2010-2019.}
		\label{abb_cycle}
	\end{figure}

	In order to quantitatively compare the annual variation of the determined number of fronts with the three methods and to avoid bias due to orography, only a box (shown in Fig. \ref{abb_na-clim}) over the North Atlantic is used.	
    Fig. \ref{abb_cycle} shows the annual cycle of the detected fronts at 600 hPa over this North Atlantic box with the TFP method, the F diagnostic and the DSI method for the three percentiles 95 \%, 97 \%, and 99 \%, respectively. Here, ten years from 2010 to 2019 (each day 0 UTC) were considered. 
    All methods detect more fronts in winter than summer, which can be explained by the northward shift of the storm tracks, while the considered box remains fixed, and, regardless of the area considered, baroclinicity, mean wind speed and thus relative vorticity are on average weaker in summer than in winter \citep[e.g.][]{Hoskins2002}. 
    With the TFP method, fewer fronts are detected each month than with the F diagnostic, which is in agreement with the global climatology previously considered.
    The range between the minimum and maximum monthly front number is greater with the F diagnostic than with all other methods, which can be explained by the fact that both baroclinicity and relative vorticity are higher on average in winter than in summer, and using their product in the F diagnostic then overemphasizes this difference. 
	The DSI method also reflects the annual cycle for all percentile thresholds, although the summer-winter spread is not as pronounced with the DSI method, especially when the threshold is increased. The fact that more fronts are recorded with the DSI method in summer than with the other methods is possibly due to the construction of the DSI. Regardless of the amount of baroclinicity itself, fronts represent a deviation from the basic state in every season. In addition, the availability of moisture for diabatic processes is larger in summer due to higher temperatures, which correlates with $\vert DSI \vert$ \citep{Claussnitzer2008,Claussnitzer2009}.	
	
	\subsection{Properties of the detected fronts} 
\begin{figure}[H]
		\centering
		\includegraphics[width=14cm]{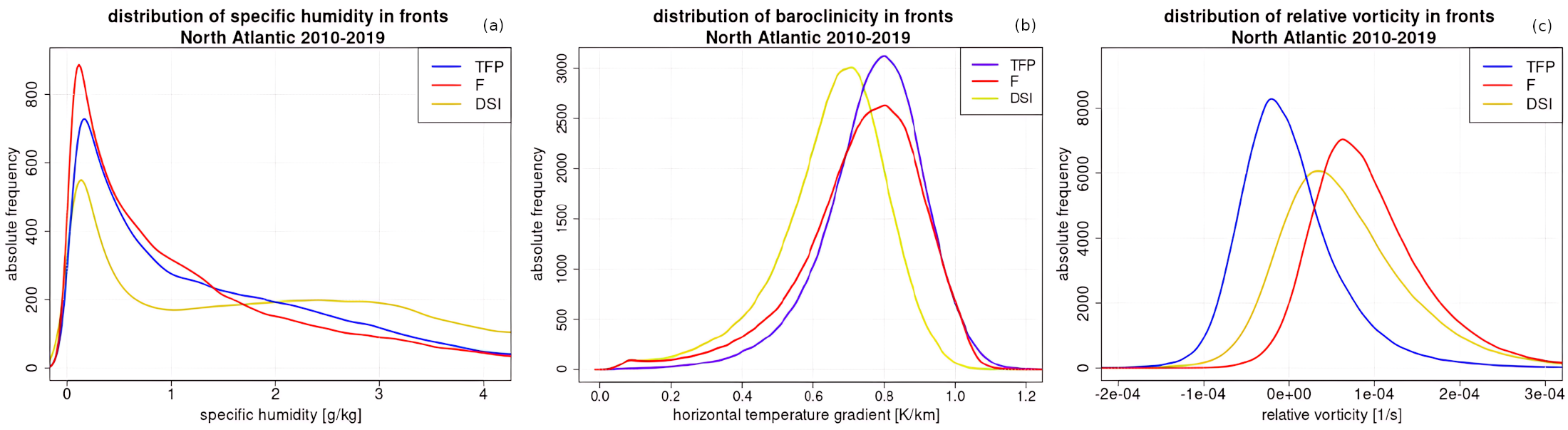}
		\caption{Front properties: Distribution of (a) specific humidity, (b) baroclinicity and (c) relative vorticity in the respective detected fronts (TFP method, F diagnostic, DSI method with 95 \% percentile) at 600 hPa over the North Atlantic between 2010-2019.}
		\label{abb_distribution}
	\end{figure}

	In this section, the properties of the detected fronts are compared. Fig. \ref{abb_distribution} shows the distribution of specific humidity, baroclinicity measured by $\Vert \nabla_h T \Vert$ and relative vorticity of the fronts detected with the TFP method, the F diagnostic and the DSI method (95 \% percentile), respectively. 
    Fronts detected with the TFP method show high baroclinicity (due to the definition of the TFP) and anticyclonic vorticity on average. The latter is not reasonable as fronts are associated with lows and itself are areas of high relative vorticity. 
    Fronts detected with the F diagnostic are characterized by high baroclinicity and high relative vorticity (both due to the definition of F) but low specific humidity. This indicates that the constituting variables in the front detection method also dominate the characteristics of the detected fronts.
    Fronts detected with the DSI method are characterized by high specific humidity and less baroclinicity. Generally, the $\vert DSI \vert$ values correlate with specific humidity, even if the specific humidity is not considered in the (dry) DSI \citep{Claussnitzer2008}. \citet{Hittmeir2021} confirmed this theoretically by deriving a hierarchy of DSI variants by stepwise including water vapor ($q_v$) and liquid water content ($q_c$) and the associated phase changes into the basic state of the DSI, which results in additional terms in Eq. \ref{gl_dsi-def}-\ref{gl_dsi-advektion} that contain the respective gradients $\nabla q_v$ and $\nabla q_c$. This emphasizes that the specific humidity is a crucial component of the diabatic processes that cause deviations from a dry basic state. The fact that fronts detected using the DSI method have less baroclinicity than fronts determined using the other methods, which contain baroclinicity directly as a factor, can be seen from the determinant form of the DSI (Eq. \ref{gl_dsi-def}): Of the six contributing terms, only four contain horizontal derivatives of potential temperature, while the static stability in the other two terms is scaled with the horizontal gradients of $\Pi,B$ that are larger than their vertical ones. Hence, the DSI reflects not only individual selected properties of fronts but their versatility, and the relative contribution of baroclinicity is thus smaller. Altogether, the DSI can be interpreted as an \textit{activation parameter}, indicating regions where energy is released, while baroclinicity is an \textit{availability parameter} describing a development potential \citep[e.g.][]{Schartner2009}.
    
	Following \citet{Hewson1998}, the movement speed of front lines, determined as contours of a quantity $\Psi$, can be estimated by 
	\begin{align} \label{gl_propagation}
	v_f = \boldsymbol{v}_h \cdot \frac{\nabla_h \Psi}{\Vert \nabla_h \Psi \Vert} = \boldsymbol{v}_h \cdot \boldsymbol{e}_{\Psi} \le \Vert \boldsymbol{v}_h \Vert.
	\end{align}
	The horizontal wind $\boldsymbol{v}_h$ is projected onto the horizontal unit vector $\boldsymbol{e}_\Psi$, so that the maximum front speed is constrained by $\Vert \boldsymbol{v}_h\Vert$. Since Eq. \ref{gl_propagation} considers the displacement of contours (and not surfaces), the front speed is calculated only for the TFP method ($\Psi =TFP$) and the DSI method ($\Psi = DSI$). The average front speed in the North Atlantic box using the TFP method is  12,6 m/s in winter and 9,5 m/s in summer, while using the DSI method it is 17,1 m/s in winter and  13,8 m/s in summer (both normalized with the respective number of fronts detected). Since the selected quantity $\Psi$ enters the calculation of the speed of movement only via the unit vector (i.e. as a direction and not as magnitude), it can be deduced that $\boldsymbol{e}_{DSI}$ is on average ''more parallel'' to $\boldsymbol{v}_h$ than $\boldsymbol{e}_{TFP}$. This confirms the capacity of the DSI to indicate the direction of movement (see, Fig. \ref{abb_dsi-pv}).


\section{Summary and discussion} \label{chap:summary}
We discuss various methods for identifying fronts from reanalyses and present a new front detection method based on the dynamic state index (DSI), which accurately measures the deviation from an adiabatic, steady and inviscid basic state that represents a general non-linear solution of the primitive equations. \\
The discussed methods and the constituting variables are summarized in Tab. \ref{tab_summary}. The standard method for front detection, which uses the TFP as purely thermal quantity, was extended by \citet{Parfitt2017} by including relative vorticity to account for the dynamical processes occurring at fronts. We point out that the F diagnostic contains the Rossby number, and thus detects fronts based on their spatial scale weighted with baroclinicity. By using the DSI, we present an alternative view on front detection by introducing a front detection algorithm that considers fronts as deviation from an adiabatic, steady and inviscid basic state. The DSI has as constituting variables potential temperature, Bernoulli function and potential vorticity and shows a dipole structure along fronts. We show that this DSI method captures the global storm track regions and reflects the annual cycle in front (and cyclone) frequency.\\
A comparison of the three methods reveals that the F diagnostic detects more fronts in winter than the TFP and DSI method. Further, the spread between summer and winter is largest in the F diagnostic, which can be explained by its construction as multiplication of baroclinicity and relative vorticity, which are positively correlated due to the transverse circulation at fronts \citep{Hoskins1982}. The DSI methods shows the smallest summer-winter spread in the number of fronts, which can be traced back to the fact that fronts represent a deviation from the basic state regardless of the season. By comparing the properties of the detected fronts, we find that fronts identified using the TFP method are characterized by high baroclinicity but on average anticyclonic relative vorticity. Fronts identified with the F diagnostic are characterized by high baroclinicity and high relative vorticity, while fronts identified with the DSI method show particularly high specific humidity. This emphasizes the role of specific humidity in diabatic processes which represent a deviation from a dry and reversible basic state. \\
Overall, the results indicate that the properties of the detected fronts strongly depend on the method used and in particular the constituting variables, which is consistent with previous findings \citep{Hope2014,Schemm2015,Spensberger2018,Soster2022}. This indicates that the front detection method should be chosen with great care and derived properties should be tested for robustness with respect to the method used. Our newly introduced DSI method can hereby provide a more general view on front detection, due to its theoretical foundation and the versatile contributing quantities and processes. The correlation of the DSI with specific humidity and precipitation sums also enables to recognize fronts based on their direct weather impact and could be particularly suitable for investigating extreme precipitation events.

	\begin{table}
		\centering
		\begin{tabular}{cccclll}
			method & quantities & frontal zones & frontal lines & properties of detected fronts\\
			\hline
			TFP & $\theta$  & yes  & yes &  baroclinicity\\
			F &$T, \: \zeta$ & yes & no&  baroclinicity, vorticity \\
			DSI & $\theta, B, \Pi$  & yes & yes &specific humidity\\
		\end{tabular}
	\caption{Characteristics of the compared front identification methods: Constituting quantities, features of the methods and properties of the respective detected fronts.}
	\label{tab_summary}
	\end{table}


\backmatter
\bmhead{Acknowledgements}
 This research has been supported by Deutsche Forschungsgemeinschaft (DFG) through grant CRC 1114 ''Scaling Cascades in Complex Systems'', Project Number 235221301, Project A01 ''Coupling a multiscale stochastic precipitation model to large scale atmospheric flow dynamics''. ECMWF is acknowledged for providing the ERA5 reanalysis data. The R open-source software package \citep{R2020} has been used to produce the analyses and graphics for this study.

\backmatter
\bmhead{Data Availability}
The used ERA5 reanalysis data is available at Copernicus Data Store (\url{https://cds.climate.copernicus.eu/#!/home}).

\backmatter
\bmhead{Code Availability} The code has been made available in form of the R Package "meteoEVT" \citep{Mack2022} (\url{https://CRAN.R-project.org/package=meteoEVT}).

\section*{Declarations}
\backmatter
\bmhead{Conflict Interest} The authors have no relevant financial or non-financial interests to disclose.

\appendix
\section{Derivation of the steady wind representation} \label{appendix:steady-wind}
	The starting point for deriving the DSI is the derivation of a general non-linear solution of the primitive equations by incorporating the basic state, i.e. stationary, inviscid and adiabatic conditions. The Navier-Stokes equation on a rotating Earth with friction under dry conditions is given by
	\begin{equation}
		\frac{d \boldsymbol{v}}{dt} + 2 \boldsymbol{\omega} \times  \boldsymbol{v} = -\frac{1}{\rho}\nabla p - \nabla\Phi + \boldsymbol{F}_R,
	\end{equation}
	where $\boldsymbol{v} = (u,v,w)^T$ represents the 3D velocity field, $\boldsymbol{\omega}$ the Earth's rotation vector, $\rho$ the air density, $p$ the pressure, $\Phi$ the gravity potential, and $\boldsymbol{F}_R$ non-conservative frictional forces.
	The total derivative of the velocity is decomposed into a stationary and advective part by using the Euler decomposition. The advective part can in turn be decomposed into a vortical and an energetic part with the Weber transformation
	\begin{equation}\label{gl_webertrafo}
	\frac{d\boldsymbol{v}}{dt} = \frac{\partial \boldsymbol{v}}{\partial t} + \boldsymbol{v}\cdot \nabla \boldsymbol{v} = \frac{\partial \boldsymbol{v}}{\partial t} + \nabla \frac{1}{2} \boldsymbol{v}^2+ \boldsymbol{\xi} \times\boldsymbol{v}.
	\end{equation}
	Using the first law of thermodynamics in the operator form $dh = Tds + \nu dp$ (with $h$ the specific enthalpy, $s$ the specific entropy and  $\nu := \rho^{-1}$) and switching the operator $d $ to $ \nabla$, an alternative form of the Navier-Stokes equations (constraint by energy conservation) can be derived
	\begin{equation} \label{ns_td}
	\frac{\partial \boldsymbol{v}}{\partial t} + (\underbrace{\boldsymbol{\xi} + 2 \boldsymbol\omega}_{=:\boldsymbol{\xi_a}}) \times \boldsymbol{v} = T\nabla s - \nabla \underbrace{ \left(\Phi + \frac{1}{2}\boldsymbol{v}^2+h\right)}_{=:B} + \boldsymbol{F}_R.
	\end{equation}
	The bracketed part is called Bernoulli function $B$ and describes the most general stream function.
	Multiplying both sides of the equation with $\times \nabla \theta$ (where $\theta$ represents the potential temperature) and using the definition of the potential vorticity  $\Pi$ leads to
	\begin{equation}
	\rho \Pi \boldsymbol{v} - \boldsymbol{\xi_a} \textcolor{darkred}{ \frac{d\theta}{dt}} +\boldsymbol{\xi_a} \textcolor{darkblue}{  \frac{\partial \theta}{\partial t} }  - \textcolor{darkgreen}{\boldsymbol{F}_R } \times \nabla \theta = - \left( \textcolor{darkblue}{ \frac{\partial\boldsymbol{v}}{\partial t}} + \nabla B \right)  \times\nabla \theta.
	\end{equation}
	
	At this point, the basic state is incorporated so that the colored terms are omitted, because the flow in the basic state is \textcolor{darkblue}{stationary}, \textcolor{darkred}{adiabatic} and \textcolor{darkgreen}{frictionless}. This yields an expression for the convective flux of the PV \citep{Schaer1993}
	\begin{equation}\label{gl_pvflux}
	\rho \Pi \boldsymbol{v} = \nabla \theta \times \nabla B.
	\end{equation}
	For non-vanishing PV, this equation can be rearranged to
	\begin{equation}
	\boldsymbol{v}_{s} = \frac{1}{\rho \Pi} \nabla \theta \times \nabla B, 
	\end{equation}
	what is called the steady-state wind representation by \citet{Schaer1993} and \citet{Nevir2004}.

\bibliography{sn-bibliography}

\end{document}